\documentclass[fleqn,10pt]{wlscirep}
\usepackage{graphicx}
\usepackage{varioref}
\usepackage{xcolor}
\usepackage{nicefrac}
\usepackage{xfrac}
\usepackage{hyperref}
\usepackage[version=4]{mhchem}
\usepackage[utf8]{inputenc}

\newcommand{\irptte}{\ce{Ir_{1-x}Pt_xTe2}}
\newcommand{\ptfour}{$\mathrm{Ir_{0.96}Pt_{0.04}Te_{2}}$}
\newcommand{\ptfive}{$\mathrm{Ir_{0.95}Pt_{0.05}Te_{2}}$}

\newcommand{\onlinecite}[1]{\hspace{-1 ex} \nocite{#1}\citenum{#1}} 

\begin{document} 
\title{Charge-Stripe Order and Superconductivity in \irptte}

\author[1,+]{O. Ivashko}
\author[2,3,+,*]{L. Yang}
\author[1]{D. Destraz}
\author[2]{E. Martino}
\author[4]{Y. Chen}
\author[4]{C. Y. Guo}
\author[4]{H. Q. Yuan}
\author[2]{A. Pisoni}
\author[2]{P. Matus}
\author[5]{S. Pyon}
\author[6]{K. Kudo}
\author[6]{M. Nohara}
\author[2]{L. Forró}
\author[3]{H. M. Rønnow}
\author[7]{M. Hücker}
\author[8]{M.~v.~Zimmermann}
\author[1]{J. Chang}

\affil[1]{Physik-Institut, Universität Zürich, Winterthurerstrasse 190, CH-8057 Zürich, Switzerland}
\affil[2]{Laboratory of Physics of Complex Matter, Institute of Physics, Ecole Polytechnique Féderale de Lausanne (EPFL), CH-1015 Lausanne, Switzerland}
\affil[3]{Laboratory for Quantum Magnetism, Institute of Physics , Ecole Polytechnique Féderale de Lausanne (EPFL), CH-1015 Lausanne, Switzerland}
\affil[4]{Center for Correlated Matter and Department of Physics, Zhejiang University, 310027 Hangzhou, Zhejiang, People's Republic of China}
\affil[5]{Department of Applied Physics, The University of Tokyo, Tokyo 113-8656, Japan}
\affil[6]{Research Institute for Interdisciplinary Science, Okayama University, Okayama 700-8530, Japan}
\affil[7]{Department of Condensed Matter Physics, Weizmann Institute of Science, Rehovot 7610001, Israel}
\affil[8]{Deutsches Elektronen-Synchrotron DESY, 22603 Hamburg, Germany}
\affil[+]{These authors contributed equally to this work}
\affil[*]{Corresponding author: linyangphy@gmail.com}

\begin{abstract}
A combined resistivity and hard x-ray diffraction study of superconductivity and 
charge ordering in Ir$_{1-x}$Pt$_x$Te$_2$, as a function of Pt substitution and 
externally applied hydrostatic pressure, is presented. Experiments are focused on 
samples near the critical composition $x_c\sim 0.045$ where competition and 
switching between charge order and superconductivity is established. We show 
that charge order as a function of pressure in \ptfive\ is preempted --- and 
hence triggered --- by a structural transition. Charge ordering appears 
uniaxially along the short crystallographic (1,0,1) domain axis with a 
(\nicefrac{1}{5},0,\nicefrac{1}{5}) modulation. Based on these results we draw a 
charge-order phase diagram and discuss the relation between stripe ordering and
superconductivity.
\end{abstract}

\maketitle

\section*{Introduction}
Transition-metal dichalcogenides have long been the centre of considerable 
attention because of their complex quasi two-dimensional electronic 
properties. Semiconductor physics~\cite{RileyNatPhys2014}, 
superconductivity~\cite{MorosanNatPhys2006,KissNatPhys2007,CostanzoNatNano2016} 
and spontaneous breaking of lattice symmetry, driven by charge-density waves 
(CDW)~\cite{WilsonAP1075,MonctonPRB1977,CastroPRL2001}, are commonly reported. 
Often, the ground state properties of these materials can be controlled by  
external non-thermal parameters such as chemical 
substitution~\cite{WagnerPRB2008}, magnetic 
field~\cite{WangPRL2016,SchmidtPRL2016} or hydrostatic 
pressure~\cite{SiposNatPhys2008}. The prototypical 1T-TaS$_2$ compound  can, for 
example, be tuned from a CDW state to superconductivity by application of 
hydrostatic pressure~\cite{SiposNatPhys2008}. Recently, a connection between 
charge density wave order in 1T-TaS$_2$ and orbital textures has been 
demonstrated~\cite{RitschelNatPhys15}. A parallel effort has been to study 
dichalcogenide systems in which spin-orbit coupling is considerable. To this 
end, IrTe$_2$ has attracted interest because spin-orbit coupling on the Ir 
site  is known to be large~\cite{KimPRL2008,SalaPRB2014}. The IrTe$_2$ system 
displays high-temperature charge ordering, and superconductivity can be induced 
by Pt or Pd substitution that in turn quenches the charge 
order~\cite{Pyon12JPSJ,Yang12PRL,Pyon13PhysicaC}. Several studies  
concluded in favour of a conventional $s$-wave pairing 
symmetry~\cite{Zhou13EPL,YuPRB2014}.  It remains however to be understood 
how  charge order, lattice symmetry and superconductivity interfere. 

In the parent compound IrTe$_2$, charge order  coincides with a  
lowering of the crystal structure symmetry (from hexagonal P$\overline{3}m1$ to 
monoclinic C$2/m$)~\cite{Pyon12JPSJ}. This effect is most likely not accidental 
and hence  IrTe$_2$ falls into the category of materials 
such as La$_{2-x}$Ba$_x$CuO$_4$\cite{HueckerPRB2011}, Ca$_2$RuO$_4$\cite{NakatsujiPRL2004,SutterNatComm2017} 
and URu$_2$Si$_2$\cite{TonegawaNatComm2014} where structural and electronic transitions appear simultaneously. 
For such systems, it is important to address 
the question whether the transition is lattice or electron
driven. Resolving this issue, is often crucial 
to understand the electronic instability.
The fact that superconductivity emerges when charge order is 
quenched by chemical pressure tuning, is probably also not coincidental. It may 
indicate that quantum criticality enters as a supporting ingredient to the 
formation of superconductivity. The interplay between charge ordering and 
superconductivity is therefore an interesting topic to explore.
Charge ordering of the parent compound has been studied in great detail, and it 
has been shown how different modulation vectors emerge as a function of 
temperature. Upon cooling the system first develops a 
(\nicefrac{1}{5},0,\nicefrac{1}{5}) modulation ($T<280$~K) that switches to 
(\nicefrac{1}{8},0,\nicefrac{1}{8}) at lower 
temperatures~\cite{Pascut14PRL,Pascut14PRB,Ko15NatComm} ($T<200$~K). There 
exist, however, no x-ray diffraction studies of the charge order in 
Ir$_{1-x}$Pt$_x$Te$_2$ near the critical composition ($x_c\sim0.045$) for 
superconductivity.

\begin{figure*}
	\centering
	\includegraphics[width=0.699\textwidth]{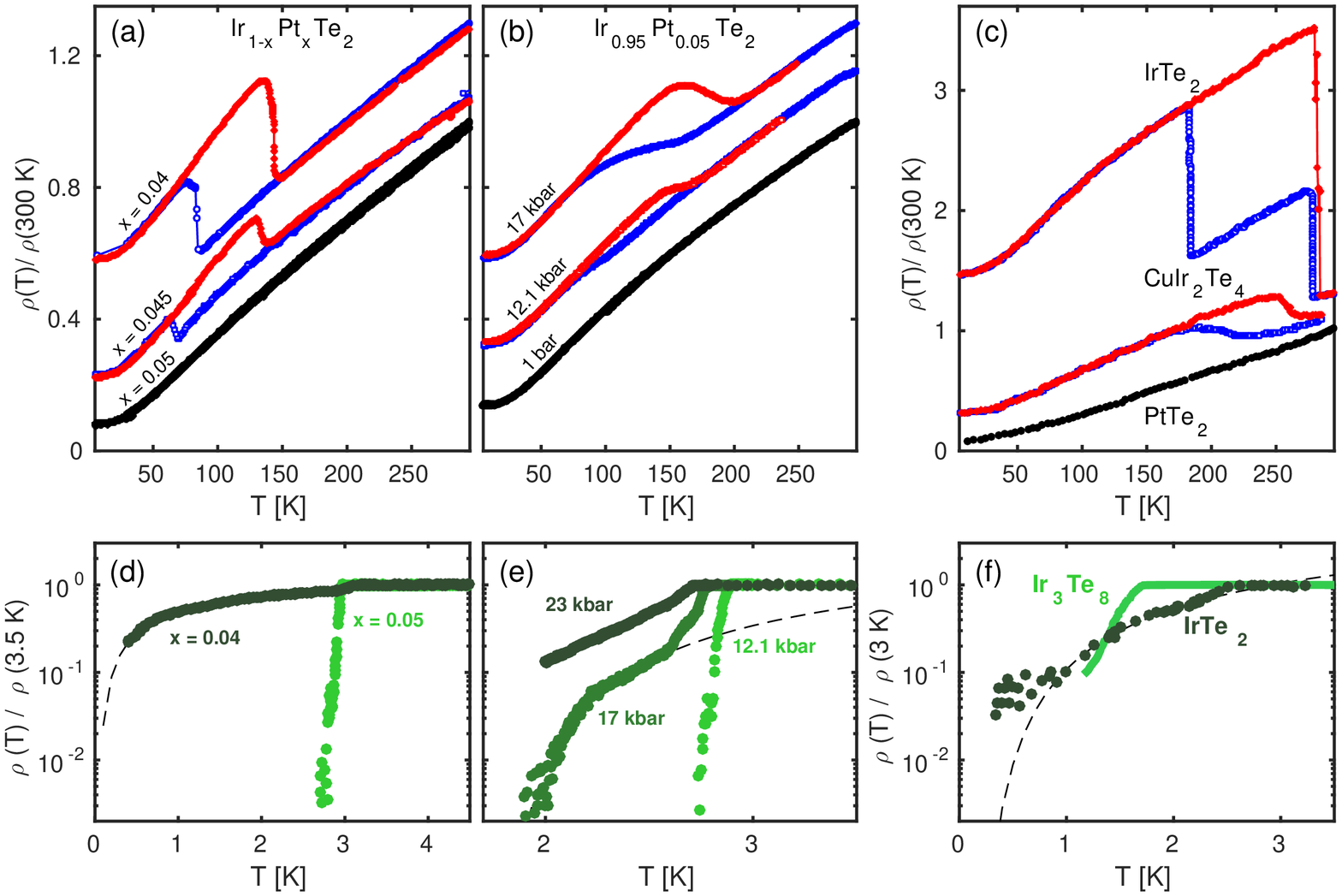}
	\caption{
	Warming and cooling resistivity curves for  \irptte\ and related 
stoichiometric compounds. (a) Substitution dependence for Pt concentrations as 
indicated. (b) Resistivity measured on \ptfive\ and hydrostatic pressures as 
indicated. (c) Resistivity curves for the parent compound IrTe$_2$, and related 
materials CuIr$_2$Te$_4$ and PtTe$_2$ (adapted from 
Ref.~\onlinecite{Matsumoto99JLTP,Ko15NatComm}).  For the sake of visibility, the 
colored curves in (a), (b) and (c) have been given an arbitrary shift. (d) and (e) 
display the low-temperature resistivity curves recorded under the same 
conditions as in (a) and (b). (f) Comparable resistivity curves of the 
stoichiometric compounds IrTe$_2$ and Ir$_3$Te$_8$ adapted from 
Refs.~\onlinecite{Fang13SciRep,LiPRB2013}. Dashed lines in (d)-(f) are guides to the eye only.}\label{Fig1}
\end{figure*}

Here we present a combined resistivity and x-ray diffraction study of \irptte\ 
as a function of chemical substitution and hydrostatic pressure near the 
critical composition $x_c$. Just below this critical composition, we find a 
temperature independent charge ordering modulation vector  
(\nicefrac{1}{5},0,\nicefrac{1}{5}). This signifies a difference from the parent 
compound where the ground state charge modulation is  
(\nicefrac{1}{8},0,\nicefrac{1}{8})~\cite{Pascut14PRB,Ko15NatComm}. 
Our pressure experiments were carried out just above $x_c$ (namely at $x=0.05$) 
in a compound with a superconducting ground state and no evidence of charge 
order at, and around, ambient conditions $1-400$~bar. With increasing pressure, we find a lowering of 
lattice symmetry above $p_{c1}\sim11.5$~kbar. This breaking of the hexagonal 
lattice symmetry appears without any trace of charge ordering that emerges only 
for pressures above $p_{c2}\sim16$~kbar. From this observation we conclude that 
charge ordering is lattice -- rather than electronically -- driven. Combining 
our results with those previously obtained in IrTe$_2$, we propose a charge 
order phase diagram as a function of Pt substitution and hydrostatic pressure. 
In terms of structure, we demonstrate that charge ordering is appearing 
unidirectionally along the short lattice parameter axis.  Finally, we discuss the 
interplay between charge ordering and superconductivity. The temperature versus 
Pt substitution phase diagram~\cite{Pyon12JPSJ} suggests that these two phases 
are competing. Based on our resistivity data, we argue that superconductivity 
may survive into the uniaxial charge ordering phase however the transition gradually broadens 
to a point where zero resistance is not observed. We discuss possible explanations 
of this effect in terms 
of (1) chemical and electronic inhomogeneity, (2) granular superconductivity
and (3) a three- to two-dimensional electronic transition.

\section*{Results}
Cooling and warming resistivity curves are plotted in Fig.~\ref{Fig1}, for 
different compositions of \irptte\ as indicated. Similar curves are shown for 
\ptfive\ for different levels of hydrostatic pressures as indicated. The hysteresis loops 
indicate a first order transition that certainly is related to the lowering of 
crystal lattice symmetry and/or the emergence of charge order. From the 
resistivity curves, alone, it is however not possible to determine whether the 
transition is electronic or lattice driven. To illustrate this point, we show in 
Fig.~\ref{Fig1}(c) resistivity curves of the stoichiometric compounds IrTe$_2$, 
CuIr$_2$Te$_4$ and PtTe$_2$. Among these materials, charge ordering has only been 
observed in IrTe$_2$. The hysteretic resistive behaviour of CuIr$_2$Te$_4$ is 
therefore not caused by charge ordering, but rather by a structural transition. In  Figs.~\ref{Fig1}(d) and (e) the 
superconducting transition of \irptte\ is displayed and compared to the 
stoichiometric compounds IrTe$_2$ and Ir$_3$Te$_8$ [Fig.~\ref{Fig1}(f)]. 
Empirically, it seems that the superconducting transition broadens dramatically in the coexistent regime.

\begin{figure*}
	\centering
	\includegraphics[width=0.999\textwidth]{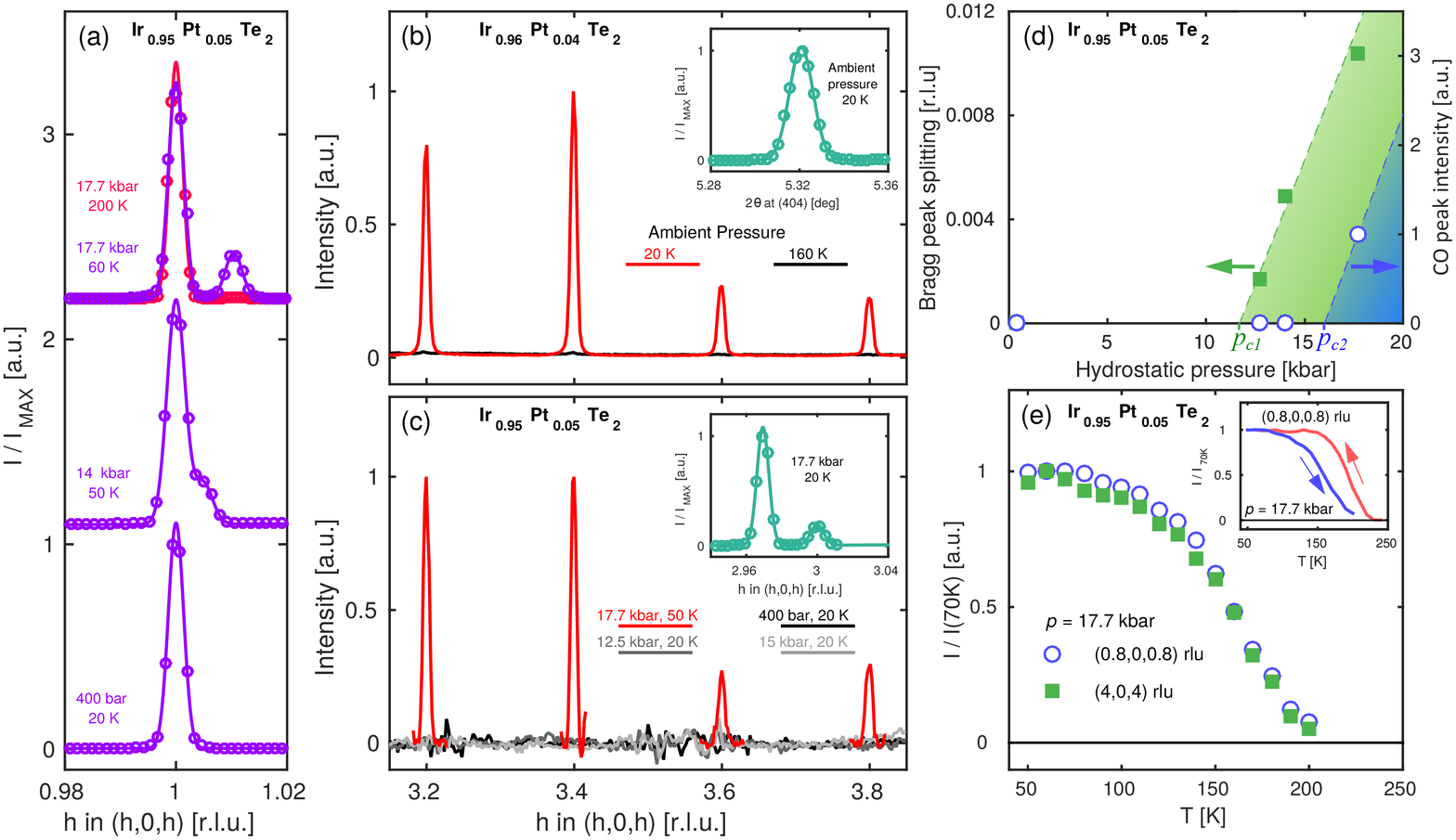}
	\caption{
	Lattice and charge ordering reflections in \irptte. (a) Bragg peak 
(1,0,1) reflection measured in \ptfive\ as a function of pressure as indicated. 
Solid lines are Gaussian fits to the data. (b) Ambient pressure x-ray diffracted 
intensity measured on \ptfour\ along the (1,0,1) direction for 20 K (red line) 
and 160 K (black line) respectively. (c) Scan as in (b) but measured at base 
temperature (20 K) on \ptfive\ for pressures as indicated. The slightly worse 
signal-to-noise level stems from the necessary background subtraction of signal 
originating from the pressure cell. (d) Bragg peak splitting and charge ordering 
intensity -- shown in (a) and (c) -- as a function of pressure. (e) Temperature 
dependence of the intensity of charge ordering and short-axis reflections on 
\ptfive with maximum applied pressure, as indicated. Warming and 
cooling intensities of  charge ordering are shown in the inset.}\label{Fig2}
\end{figure*}

To gain further insight into the relation between the lattice and charge order, 
we carried out an x-ray diffraction study. In Fig.~\ref{Fig2}(a), we show the 
fundamental lattice Bragg peak $\boldsymbol{\tau}=(1,0,1)$ measured at low 
temperature on \ptfive\ at different pressures as indicated. At low pressure ($p=400$~bar)
a single sharp Bragg peak is observed. Above a critical pressure $p_{c1}$, this peak 
develops a shoulder that upon further increased pressure evolves into a separate 
Bragg peak. When heating above 200 K, this Bragg peak splitting disappears. 
Altogether, this evidences a low-temperature pressure-induced lowering of the 
lattice symmetry.

Next, we explore the charge ordering. 
$\mathbf{Q}$-scans recorded on \ptfour\ along the $(h,0,h)$ high symmetry 
direction are displayed in Fig.~\ref{Fig2}(b). Just as reported in IrTe$_2$~\cite{Pascut14PRL,Pascut14PRB}, no 
twinning was observed on Bragg peaks equivalent to $\boldsymbol{\tau}=(1,0,1)$ 
--- see inset. Moreover, below 160 K strong charge order reflections are 
observed at wave vectors  $\mathbf{Q}=\boldsymbol{\tau}+\mathbf{q}_{co}$ where 
$\mathbf{q}_{co}=(\pm\nicefrac{1}{5},0,\pm \nicefrac{1}{5})$ and 
$(\pm\nicefrac{2}{5},0,\pm \nicefrac{2}{5})$ and $\boldsymbol{\tau}$ are 
fundamental Bragg reflections. We find (not shown) that off-diagonal reflections 
of the type $(h,0,h+n)$ with $n=1,2,3$ are much weaker than for $n=0$. As the 
diffracted intensity $I$ is proportional  to $\mathbf{Q}\cdot\mathbf{u}$ where 
$\mathbf{u}$ is the atomic displacement~\cite{ChangNatPhys2012,BlackburnPRL2013}, we conclude that displacements are 
predominately along the $(h,0,h)$ direction.

With this knowledge, we studied the charge order in the pressure-induced twinned 
phase of \ptfive. The crystal was carefully aligned on the 
$\boldsymbol{\tau}=(3,0,3)$ Bragg peak using the larger lattice constant. At the 
highest applied pressure $p\simeq17.7$~kbar, a 
$\mathbf{q}_{co}=(\pm\nicefrac{1}{5},0,\pm \nicefrac{1}{5})$ charge modulation 
is observed with respect to the Bragg peak with the shorter lattice parameter 
[see Fig.~\ref{Fig2}(c)]. The charge ordering reflection displays, just as the 
resistivity curves, hysteretic behaviour as a function of temperature [inset of 
Fig.~\ref{Fig2}(e)]. Finally, we show in Fig.~\ref{Fig2}(e) how upon cooling the 
charge order reflection and the short-axis Bragg peak $\tau=(4,0,4)$ have identical temperature 
dependence. This demonstrates an intimate relation between the crystal lattice 
symmetry breaking and charge ordering.

\begin{figure}
	\centering
	\includegraphics[width=0.699\textwidth]{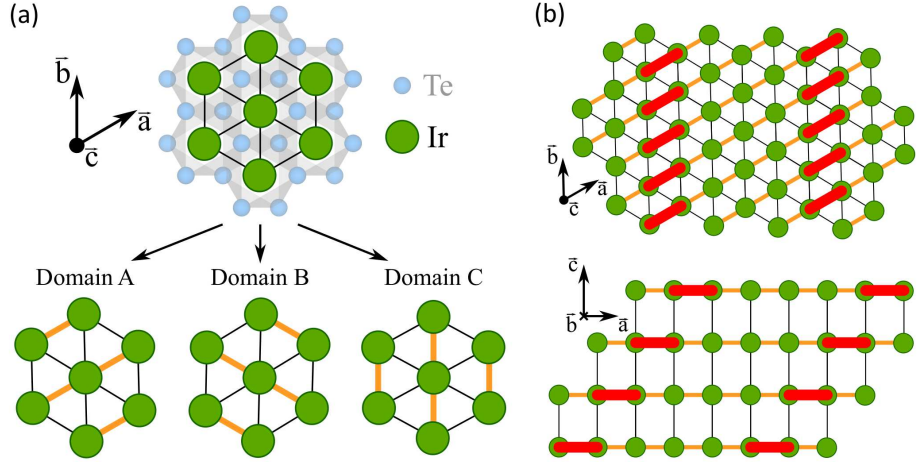}
	\caption{
	(a) Projection of the hexagonal crystal structure of 
IrTe$_2$. The transition into monoclinic structure implies 
formation of three domains where a short lattice parameter axis is found along 
the $\vec{a}$, $\vec{b}$ or $\vec{a}-\vec{b}$ direction. These domains are 
labeled A, B and C respectively. (b) Stripe charge order forms along the short 
axis direction. The Ir$^{3+}$-~Ir$^{3+}$ dimers -- indicated by red bonds -- 
intersect the crystal structures with  $\vec{b}$, $\vec{a}+\vec{c}$  
planes.}\label{Fig3}
\end{figure}

\section*{Discussion / Interpretation}
\begin{figure*}
	\centering
	\includegraphics[width=0.999\textwidth]{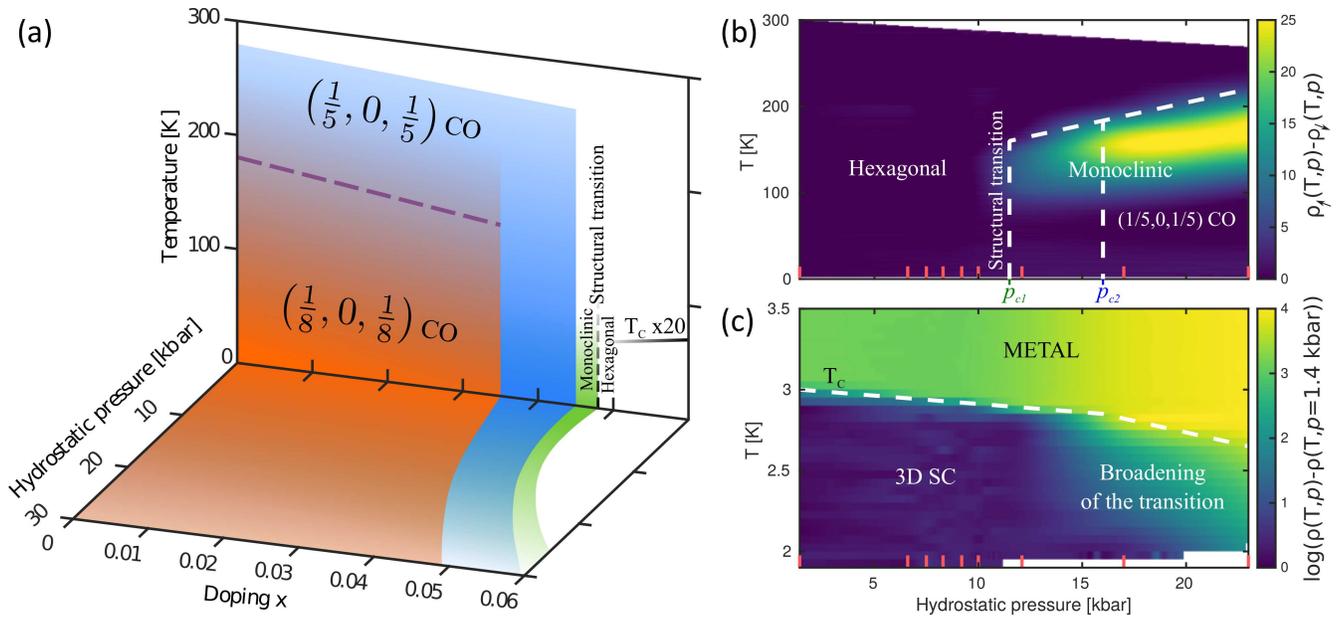}
	\caption{(a) Schematic pressure - temperature phase diagrams of the charge 
ordering and crystal lattice twinning of \irptte. (b) Hydrostatic pressure vs 
temperature map of the difference between the warming and cooling resistivity 
curves of \ptfive ~represented in false colours. (c) Similar map but for the difference of 
each resistivity curve with the one measured at $1.4$~kbar in the superconductor 
transition temperature range (displayed in logarithmic-intensity scale). Red 
ticks indicate the measured pressures. White dashed lines are guides to the 
eye.}\label{Fig4}
\end{figure*}

\textit{Lattice vs electronic mechanism:} We start by discussing the nature of 
the charge ordering transition. The pressure-induced Bragg peak splitting 
[Fig.~\ref{Fig2}(a)] is most naturally explained in terms of domain formation 
caused by a lowering of the crystal lattice symmetry. In essence, our experiment 
suggests that the lattice parameters along the (1,0,1) and (0,1,1) directions 
become inequivalent under application of pressure. The system thus develops 
three domains with a short lattice parameter along the $\vec{a}$, $\vec{b}$ or 
$\vec{a}-\vec{b}$ axes, see Fig.~\ref{Fig3}(a). All three types of domain are 
observed when scanning along the (1,0,1) direction in the pressure-induced 
twinned phase and hence two Bragg peaks are found -- shown in 
Fig.~\ref{Fig2}(a). This  twinning effect clearly appears before charge 
ordering, suggesting that the latter is lattice driven. Given 
that we observe the same (\nicefrac{1}{5},0,\nicefrac{1}{5}) modulation as in 
IrTe$_2$ (high-temperature), it is not inconceivable that the same conclusion 
applies to the parent compound. Combining our results with previous studies of 
IrTe$_2$, we propose in Fig.~\ref{Fig4}(a) a schematic pressure, Pt substitution 
and temperature phase diagram  including the charge ordering and the structural 
hexagonal to monoclinic transition.

\textit{Charge order structure:} 
The surface and bulk charge ordering structure of IrTe$_2$ has been studied by 
scanning tunnelling microscopy 
(STM)~\cite{MachidaPRB2013,LiNatCom2014,DaiPRB2014,KimSTMPRB2014,MauererPRB2016,
ChenPRB2017} and x-ray diffraction~\cite{Pascut14PRL,Pascut14PRB,TakuboPRB2014} 
techniques. The STM studies generally find uniaxial charge ordering structures. 
Furthermore, differences in charge modulations between the bulk and surface have 
been pointed out~\cite{ChenPRB2017}. Our bulk-sensitive results on \ptfive\ 
indicate that the pressure-induced charge order is connected to the short-axis 
direction only. Therefore, the most simple explanation is uniaxial 
Ir$^{3+}$-Ir$^{3+}$ dimer formation  along the short lattice parameter axis as 
illustrated in Fig.~\ref{Fig3}(b). For such a structure, an electronic gap is 
expected only along the reciprocal short lattice parameter axis. However since 
the crystals are inevitably twinned along three different directions,
it can be challenging to observe
with angle resolved photoemission spectroscopy (ARPES) 
experiments
, in particular when factoring in  
the complex electronic band structure.~\cite{OotsukiJPSJ2013,OotsukiPRB2014,Kim15PRL}
A suppression of the spectral weight (near the Fermi level) is observed 
with ARPES and optical experiments. This observation is at odd with  a conventional 
charge density wave and hence taken as evidence of novel type of 
charge ordering~\cite{OotsukiJPSJ2013,Kim15PRL,Fang13SciRep}.

\textit{Superconductivity and Charge order:}
Finally, we discuss the relation between unidirectional charge order and 
superconductivity. 
From our pressure-dependent x-ray and resistivity 
experiments, we show that a lowering of the crystal symmetry has no impact 
on superconductivity [see Fig.~\ref{Fig1}(e) and Fig.~\ref{Fig4}(b)]. 
Upon entering into the charge ordered phase, the superconducting transition however, 
broadens dramatically. While the initial superconducting onset remains fairly constant, 
the onset of zero  resistance (within the detection limit) undergoes dramatic changes.
In fact as a function of pressure, the system quickly reaches a regime where zero resistance
is not observed within the measured temperature window [see Fig.~\ref{Fig1}(e)]. The same trend
is found at ambient pressure when lowering the Pt content [see Figs.~\ref{Fig1}(d) and (f)]. 
Hence there seems to be a correlation between the occurrence of the charge order and a 
broadening of the superconducting transition. On general grounds, such a broadening can have
different explanations. (1) Chemical or electronic inhomogeneities can smear the transition.
(2) Granular superconductivity is also characterised by broad transitions. 
(3) Low-dimensional superconductivity is known 
to introduce two temperature scales. In particular, for two-dimensional superconductivity,
an exponential resistive  drop, approximately described by 
$\rho(T)\propto\exp\left(\frac{-b}{\sqrt{t}}\right)$,
is expected below $T_c$. Here $b$ is a constant and 
$t=(T-T_c^{3D})/T_c^{3D}$ with $T_c^{3D}$ being a second superconducting temperature scale. 
This Kosterlitz - Thouless transition~\cite{Li07PRL,BenfattoPRB2009} scenario finds its 
relevance in \irptte, since charge order is shown to generate 
two dimensional walls of low density-of-states.~\cite{Pascut14PRL,Pascut14PRB,TatsuyaJPSJ2014,EomPRL2014,BlakePRB2015}
It is therefore not inconceivable that superconductivity is suppressed inside these walls. 
Hence there exists a possible physical mechanism for two-dimensional superconductivity in \irptte. 
Further experimental evidence supporting this scenario would be of great interest.  
Based on the experimental  evidence presented here, it is difficult to prove the 
Kosterlitz - Thouless scenario. Nor can we completely exclude inhomogeneities or grain boundaries.
Chemical inhomogeneity is very unlikely to be the cause, since it should not be
influenced by hydrostatic pressure. Inhomogeneous pressure can also be excluded as 
the broadening is found also at ambient pressure [see Fig.~\ref{Fig1}(d)]. 
Intrinsic electronic inhomogeneity could be tuned by both pressure and chemical substitution.
However, one would expect that inhomogeneity generates more modest correlation length of
the charge order. Experimentally, however, long range (resolution limited) charge order
reflections are observed. The domain formation makes the granular superconducting scenario
more plausible. We notice, however, that the pressure induced crystal domain formation 
initially have no influence on superconductivity.
Explaining our data in terms of granular superconductivity is therefore not straightforward.

\section*{Conclusion}
In summary, we have presented a combined resistivity and x-ray diffraction study 
of \irptte\ as a function of Pt substitution and hydrostatic pressure. Just 
below the critical composition $x_c\sim0.045$ charge order with a 
(\nicefrac{1}{5},0,\nicefrac{1}{5}) wave vector is found. The same modulation 
appears in \ptfive\ upon application of hydrostatic pressures beyond 
$p_{c2}\sim16$~kbar. Based on these observations a charge ordering phase diagram 
is constructed. Application of pressure furthermore revealed a lattice symmetry 
lowering transition appearing before the charge ordering. We thus conclude that 
the charge ordering in  \irptte\ is lattice driven.
Finally, we discussed the relation between charge order and superconductivity.

\section*{Methods}
Single crystals of \irptte\ were grown using a self-flux 
technique~\cite{Fang13SciRep}. Piston-type pressure cells~\cite{Zimmermann08RCI} 
with Daphne oil as pressure medium were used to reach $\sim18$~kbar and 
$23$~kbar, for x-ray diffraction and resistivity experiments respectively. The 
hydrostatic pressure was estimated from the orthorhombicity of  
La$_{1.85}$Ba$_{0.125}$CuO$_4$ at $60$~K~\cite{Huecker10PRL} and the resistive 
superconducting transition of lead. The electrical resistivity was measured by a 
conventional four-probe method using a physical property measurement system 
(Quantum Design PPMS-14T) and hard x-ray diffraction (100 keV) experiments were 
carried out with the triple-axis instrument at beamline P07 at PETRA III, DESY. 
Although \irptte\ at certain temperatures and pressures displays crystal 
structure twinning, the momentum $\mathbf{Q}=(h,k,l)$ is presented in hexagonal 
notation with $a\approx b\approx3.95$~\AA\,~and $c\approx5.38$~\AA. 
Crystallographic projections were produced using the VESTA software~\cite{MommaJAC2011}. 

The datasets generated during and/or analysed during the current study are available 
from the corresponding author on reasonable request.

\bibliography{bibl}

\section*{Acknowledgements}
This work was supported by the Swiss National Science Foundation through its 
Sinergia network MPBH and Grant No. BSSGI0\textunderscore 155873 and PP00P2\textunderscore 150573.
Work at Zhejiang University was supported by the National Key R\&D Program of China 
(Grants No. 2016YFA0300202 and No. 2017YFA0303100) and the National Natural Science 
Foundation of China (Grant No. 11474250).

\section*{Author contributions statement}
S.P., K.K., and M.N. grew the \irptte\ crystals. L.Y., D.D., M.E., Y.C., A.P., P.M., 
H.M.R. and L.F. conducted the resistivity measurements.  O.I., M.H., M.v.Z., 
and J.C. carried out the x-ray diffraction experiments. All co-authors 
contributed to the manuscript.

\section*{Competing financial interests}
The authors declare that they have no competing interests.
\end{document}